\documentclass[twocolumn,preprintnumbers,superscriptaddress,nofootinbib,aps,prd,floatfix]{revtex4}

\usepackage{amsmath,amssymb}
\usepackage{graphicx} 
\usepackage{epstopdf} 
\usepackage{slashed}
\usepackage{subfigure}
\usepackage[usenames,dvipsnames]{color}
\usepackage{multirow,enumerate}
\usepackage{hyperref}

\newcommand{\CP}{$\mathcal{CP}$}

\newcommand{\ab}{\text{ab}}
\newcommand{\nn}{\nonumber}

\begin{document}

\title{Bounding the top Yukawa with Higgs-associated single-top production}

\author{Christoph Englert} \email{christoph.englert@glasgow.ac.uk}
\affiliation{SUPA, School of Physics and Astronomy, University of
  Glasgow,\\Glasgow, G12 8QQ, United Kingdom\\[0.2cm]}
\author{Emanuele Re} \email{emanuele.re@physics.ox.ac.uk}
\affiliation{Rudolf Peierls Centre for Theoretical Physics, Department
  of Physics,\\University of Oxford, Oxford, OX1 3NP, United Kingdom}

\begin{abstract}
  After the discovery of the 125 GeV scalar boson with gauge
  properties similar to the Standard Model Higgs, the search for
  beyond the SM interactions will focus on studying the discovered
  particles' coupling properties more precisely and shedding light on
  the relation of fermion masses with the electroweak vacuum. The
  large mass of the top quark and the SM-predicted order one top
  Yukawa coupling is a natural candidate for BSM physics, though
  experimentally challenging to constrain. In this paper, we argue
  that investigating angular correlations in $pp\to tHj$ production
  provides an excellent handle to constrain the top Yukawa coupling
  $y_t$ via direct measurements, even when we focus on rare exclusive
  final states. We perform a hadron-level analysis and show that we
  may expect to constrain $y_t\gtrsim 0.5\, y_t^{\text{SM}}$ at
  95\%-99\% confidence level at the high luminosity LHC using
  semi-leptonic top decays and $H\to \gamma \gamma$ alone, by
  employing a two-channel measurement approach.
\end{abstract}

\preprint{OUTP-14-01P}

\maketitle 


\section{Introduction}
\label{sec:intro}
The discovery of a 125 GeV scalar boson~\cite{:2012gk,:2012gu} marks a
milestone in our understanding of the mechanism of electroweak (EW)
symmetry-breaking. In order to unambiguously decipher the role exactly
played by the corresponding scalar field in breaking the EW symmetry,
it is mandatory to measure as accurately as possible the couplings
between the Higgs boson and all other SM particles we already know, as
well as the Higgs self-coupling. This programme has already started
and, as new data becomes available, results are continuously updated
by the ATLAS and the CMS collaborations~\cite{newboundsa,newboundsb},
as well as by the theoretical community~\cite{theo}.

If no striking direct evidence for new physics will be found within
the first few years of the next LHC phase, an accurate extraction of
the Higgs couplings will become even more important than it is already
now: looking for deviations from the SM values will then be our main
route to probe (indirect) manifestations of new physics. In other
words, if no other new particle beside the Higgs is found, one of the
main goals in the near future will be precision physics in the Higgs
sector, using data from the LHC as well as from other experiments (see
{\it e.g.}~\cite{Brod:2013cka,michi,Contino:2013kra} for discussions).

The extraction of the Higgs mass, quantum numbers and couplings (and
the related confidence levels) from LHC data is usually performed by
minimizing a chi-squared distribution associated with a global fit to
the data. Although theoretically debatable, it is common practice to
choose the coefficients representing deviations from the SM values of
the Higgs couplings as free parameters in this
procedure~\cite{newboundsa,newboundsb,theo}. The results of such fits
can be used to directly constrain the parameter space of specific
extensions of the SM, or to map deviations from the SM onto the
coefficients of higher-dimensional operators, using an effective field
theory language.

The ultimate accuracy of this approach will be limited by systematics,
statistics, and theoretical uncertainties in the prediction of signal
and background cross sections and branching ratios (these are the
quantities used to define the so-called signal strengths, {\it i.e.}
the quantities used to obtain the set of Higgs couplings for which the
best fit to data is obtained).

A precise (in)direct measurement of the top Yukawa coupling $y_t$ (or
at least direct sensitivity to it) is of fundamental importance. The
large mass hierarchy between the different quark generations is not
explained in the SM and the top mass being close to the electroweak
scale can be interpreted as a hint for TeV-scale physics beyond the
SM. Well-known examples of modified Higgs-top interactions are the two
Higgs doublet model, the MSSM and composite Higgs scenarios where the
size of the top mass is explained by linear mixing effects with new
TeV-scale top
partners~\cite{Agashe:2004rs,Azatov:2013ura,Dawson:2012di}. In the
latter models, the contribution from the Higgs vacuum expectation
value is less constrained, and the top Yukawa can be smaller than the
SM value, $y_t< y_t^{\text{SM}}$.

In the light of the currently available data, the aforementioned fits
are sensitive to $y_t$ mainly via the measurement of the cross section
for Higgs production in gluon fusion as well as the Higgs to diphoton
branching ratio. Both the $gg\to H$ and $H\to \gamma\gamma$ processes
are loop-mediated, and therefore the extraction of $y_t$ from these
measurements is potentially very sensitive to effects of to yet-to-be
discovered states; large deviations from the SM expectations in these
channels would be a strong hint of new physics.  However, if the Higgs
is indeed a pseudo-Nambu Goldstone boson, the effective $ggH$ and
$\gamma\gamma H$ couplings can still be SM-like, because higher
dimension operators are suppressed by the approximate shift symmetry
of the Goldstone Higgs doublet. In such a case, a direct measurement
of the top Yukawa coupling provides valuable information necessary to
break the measurements' degeneracy in the extended top sector, where
an enlarged (global) symmetry is responsible for the ``conspiracy'' to
SM-like $ggH$ and $\gamma\gamma H$
couplings~\cite{Azatov:2013ura}. This is especially true when the top
partners fall outside the LHC coverage or are masked by experimental
systematics. Similar phenomenological implications also hold for
exotic models with Higgs triplets, see {\it
  e.g.}~\cite{Englert:2013zpa}.

The above examples clearly show that, despite being extremely
interesting and seminal to Higgs physics, the presence of potentially
unknown loop effects (in addition to the LHC being unable to directly
measure the total Higgs width to satisfactory accuracy) makes the
$ggH$ and $\gamma\gamma H$ not ideal to set theoretically solid bounds
on the tree-level $t\bar{t}H$ coupling in a model-independent way. It
is therefore important to complement these indirect measurements with
direct observations of processes where $y_t$ enters already at
tree-level.

At the LHC there are two basic processes which serve this purpose:
$t\bar{t}$ associated production ($pp\to t\bar{t}H$) and Higgs +
single-top production ($pp\to t H j$). An experimental observation of
these production channels is challenging because cross sections are in
general quite small ($t\bar{t}H$ has the smallest cross section among
the standard Higgs-production processes), and, moreover, backgrounds
are generically hard to suppress.

Not surprisingly, until new techniques were introduced few years
ago~\cite{Butterworth:2008iy,Plehn:2009rk,Soper:2012pb,Curtin:2013zua,Artoisenet:2013vfa},
there were serious doubts even about being able to observe the $pp\to
t\bar{t}H$ signal on top of the dominant backgrounds~\cite{schumacher}
in the first place. Associated single-top production~\cite{original}
has an even slightly smaller cross section, and, for similar reasons,
has received little attention~\cite{Maltoni:2001hu,Tait:2000sh} until
recently. Despite of the aforementioned experimental difficulties,
given the importance of the top Yukawa as a parameter potentially
probing new physics, it is worthwhile to investigate the signatures
that might allow its direct extraction. Although current projections
indicate that a direct measurement will be challenging (at least with
traditional analysis techniques), studying the extent to which $y_t$
can be directly bounded remains a relevant and timely question.

\begin{figure}[!t]
  \includegraphics[width=0.45\textwidth]{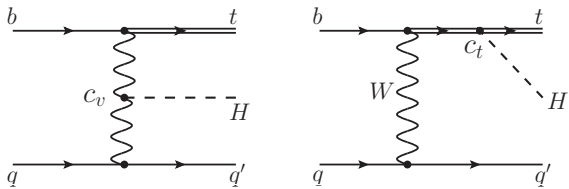}
  \caption{\label{fig:FD} Feynman diagrams for tree-level production
    of single-top + Higgs. The diagram on the left (right) is
    proportional to $C_{WWH}$ ($C_{ttH}$).}
\end{figure}

The purpose of this paper is to perform a phenomenological analysis of
a signal based on associated single-top production, and to discuss in
how far we can use a successful signal and background analysis to
constrain $y_t$. We will show that despite the fact that the top
semileptonic and the $H\to \gamma \gamma$ branchings are not the
dominant ones, it is still possible to obtain limits for the
high-luminosity LHC.

In Sec.~\ref{sec:pheno} we briefly overview the phenomenology of Higgs
+ single-top production.  In Sec.~\ref{sec:14tev} we detail our
analysis and present our results, before we summarize our findings and
conclude in Sec.~\ref{sec:conc}.

\begin{figure*}[!t]
  \includegraphics[width=0.45\textwidth]{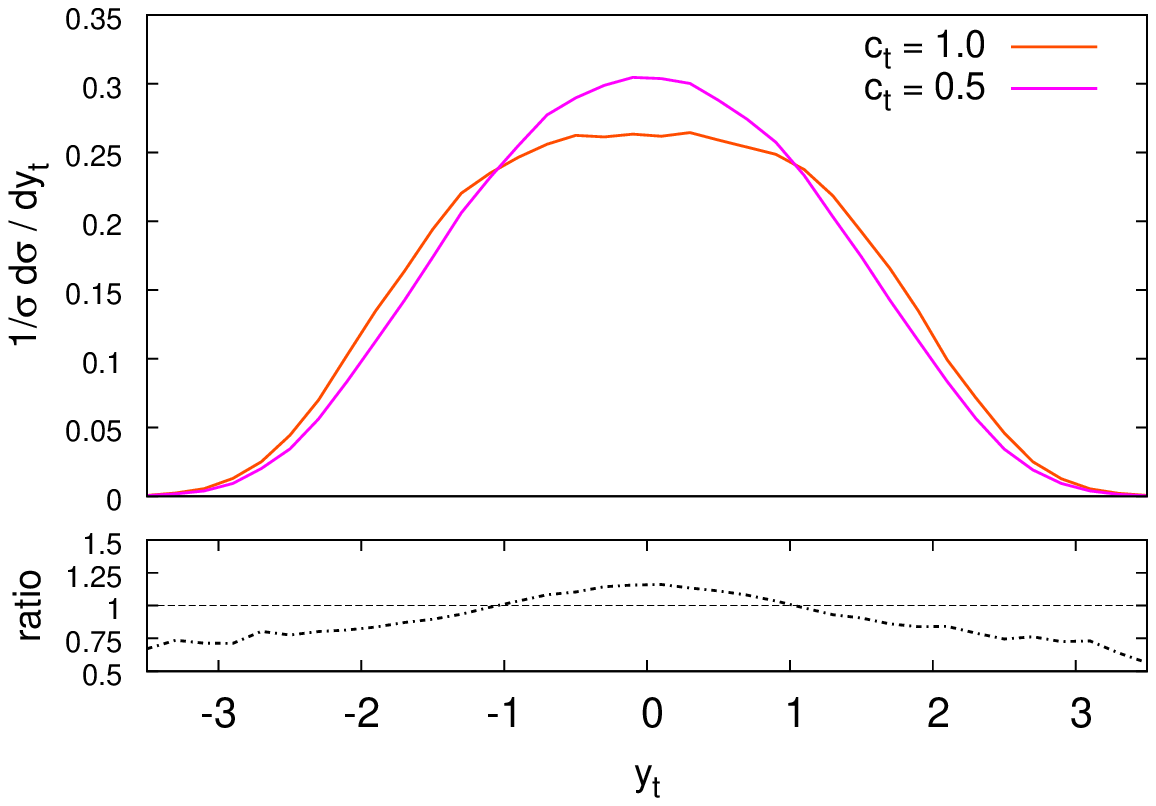}
  \hspace{0.05\textwidth}
  \includegraphics[width=0.45\textwidth]{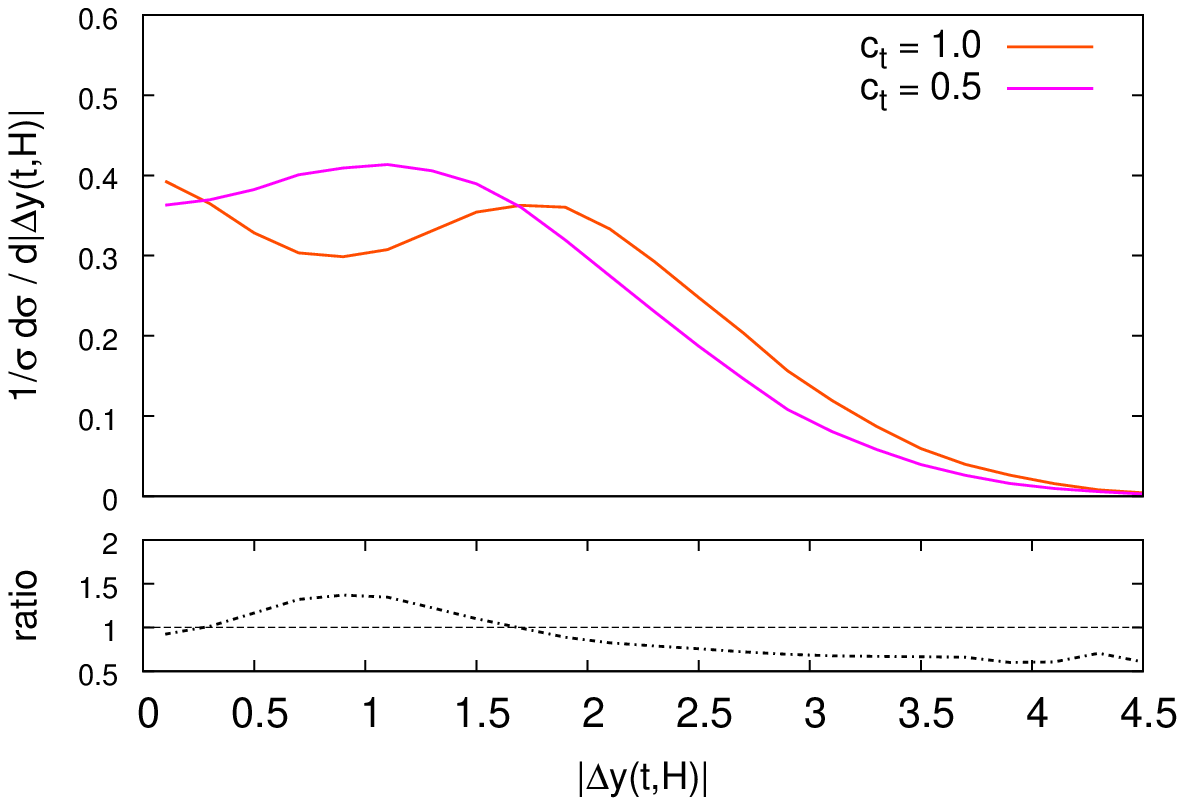}
  \caption{\label{fig:yt_dytH_LO} Parton level leading-order
    distribution of the reconstructed top quark rapidity $y_t$ (left)
    and the rapidity distance between the top quark and the Higgs
    boson $|\Delta y(t, H)|$ (right).  Plots have been obtained using
    the cuts in Eq.~\ref{eq:basiccuts} and have been normalized to
    unity. In the lower insets the ratio between the $c_t=0.5$ and the
    SM distributions is shown.}
\end{figure*}

\begin{figure*}[!t]
  \includegraphics[width=0.45\textwidth]{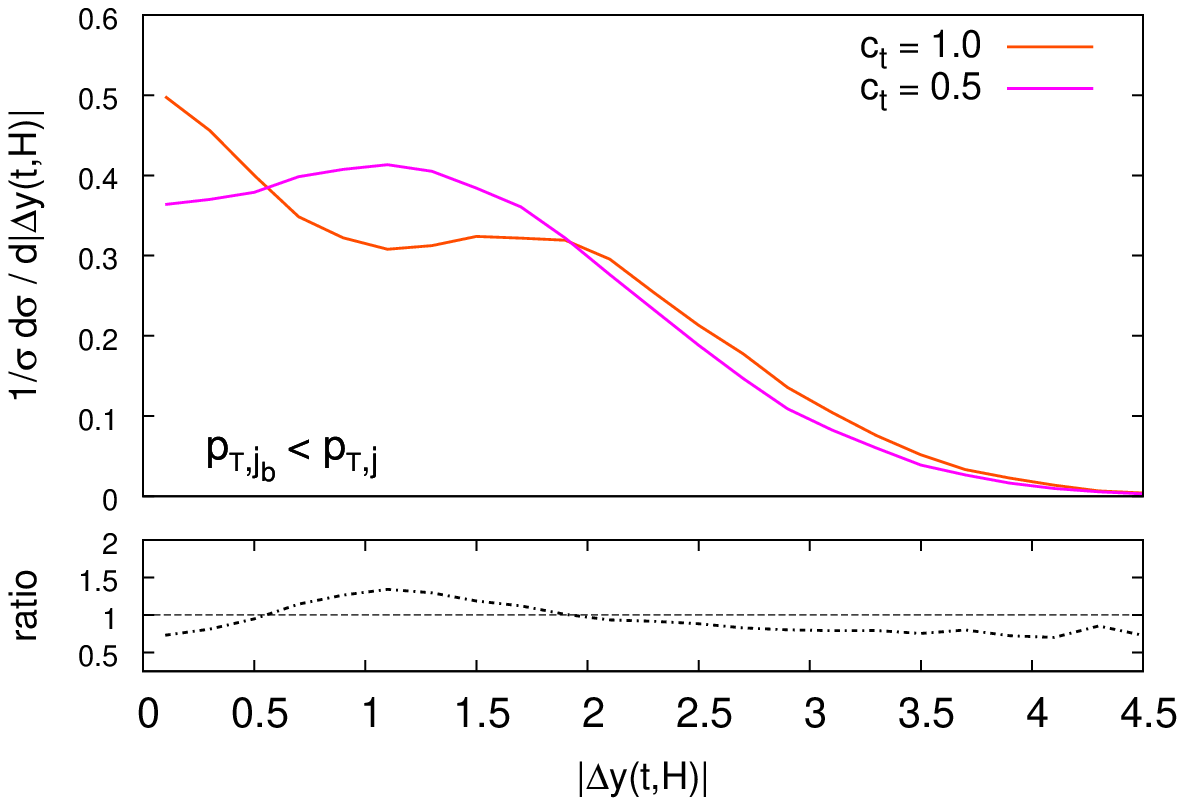}
  \hspace{0.05\textwidth}
  \includegraphics[width=0.45\textwidth]{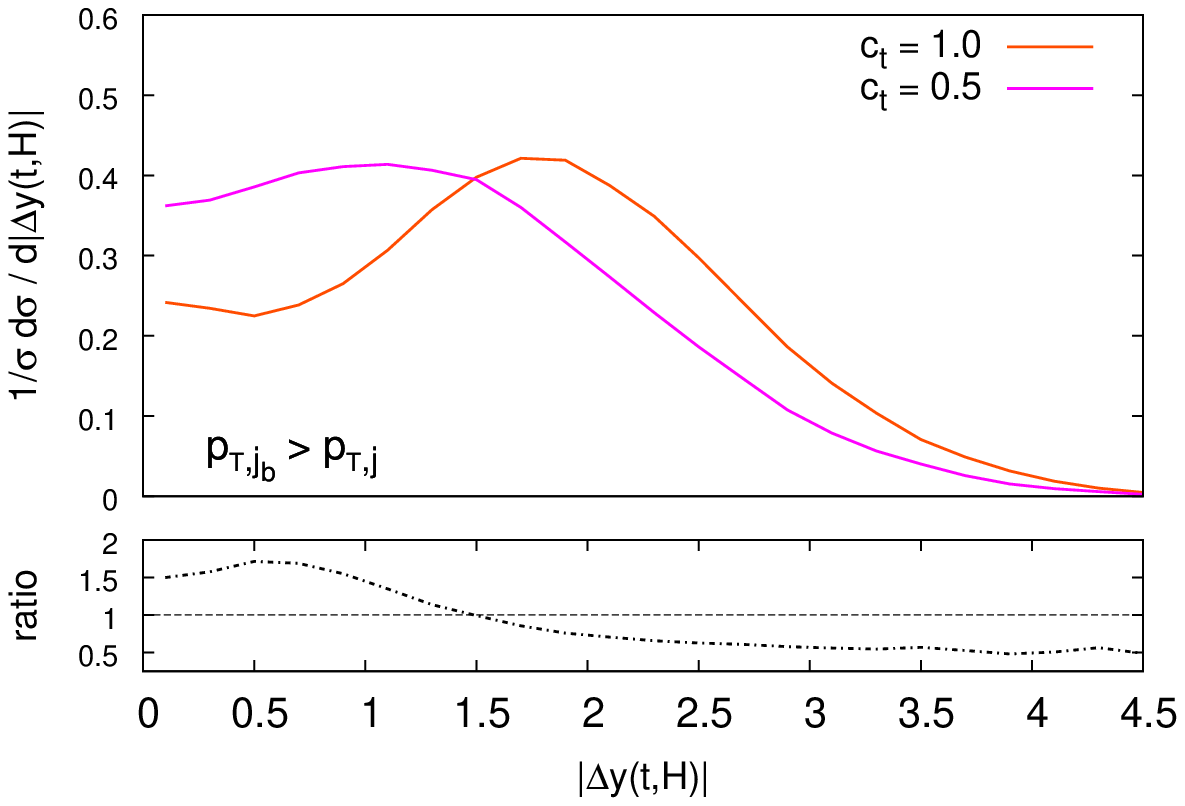}
  \caption{\label{fig:dytH_jb_bj_LO} Rapidity distance between the top
    quark and the Higgs boson in presence of the cuts in
    Eq.~\ref{eq:basiccuts}.  On the left (right) panel, the
    requirement of having a $b$-jet softer (harder) than the light jet
    is enforced.}
\end{figure*}

\section{Higgs + single-top phenomenology}
\label{sec:pheno}
At the lowest order in perturbation theory, the hadroproduction of
Higgs + single-top arises from the Feynman diagrams shown in
Fig.~\ref{fig:FD}.\footnote{In this paper we work in the 5-flavour
  scheme, which means that in the initial state we consider massless
  $b$-quark. For this reason, we neglect diagrams with the Higgs being
  emitted off $b$-quarks. In the single-top literature, it is known
  that a 5-flavour approach compares well with a computation in the
  4-flavour scheme, that would allow in turn a better description of
  the spectator
  $b$-jet~\cite{Campbell:2009ss,Maltoni:2012pa,Frederix:2012dh}.}
These two diagrams show that the top Yukawa enters at tree-level, and,
moreover, because of the interference taking place at the amplitude
level, the squared matrix-element contains a term linear in
$C_{ttH}C_{WWH}$, where we have parameterized the deviations from the
SM Higgs couplings
\begin{equation}
  C_{ttH}^{\rm SM}=-\frac{y_t}{\sqrt{2}}=-\frac{m_t}{v_{\rm SM}}\,, \
  \ \ 
  \ C_{WWH}^{\rm SM}=g_W^2\frac{v_{\rm SM}}{2}
\end{equation}
as
\begin{equation}
  C_{ttH}=c_t\times C_{ttH}^{\rm SM}\,, \ \ 
  \ \ C_{WWH}=c_v\times C_{WWH}^{\rm SM}\,.
\end{equation}
Since the Higgs insertion at a fermion line introduces a chirality
flip, dialling $C_{ttH}$ away from its SM value is tantamount to
populating different (anti-)top helicity states that in turn result in
different top decay patterns in comparison to the SM. It is the
combination of these effects that motivates the $tHj$ channel as a
unique tool for establishing a direct measurement of sign and size of
the top Yukawa as opposed to $pp\to t\bar{t}H$. Hence it is no
surprise that $tHj$ production has recently received considerable
attention from the theory
community~\cite{Biswas:2012bd,Agrawal:2012ga,Farina:2012xp,Biswas:2013xva}.

Recent global fits of the Higgs couplings are now ruling out the
$c_t<0$ possibility well above the 95\%
CL~\cite{Aad:2013wqa,Chatrchyan:2013iaa}, and start constraining the
$c_t> 0$ parameter region with available Higgs
data.\footnote{Obviously direct measurement constraints are
  statistically limited in the present case and will be not as tight
  as the indirectly obtained ones.}  Very recently it has also been
noticed that a \CP{}-violating component to the top Yukawa coupling
($\sim i\, \tilde{C}_{ttH} (\bar{t}\gamma_5 t)\, H$) can be studied in
this channel~\cite{michi}, complementing bounds obtained from
low-energy experiments~\cite{Brod:2013cka}. We do not include this
possibility in this paper, but we expect that similar implications can
be formulated in the \CP{}-violating context, too.

If also $C_{WWH}$ is assumed to be a free parameter, then deviations
from the SM expected cross section could be used to set bounds in the
$(c_v,c_t)$ plane. Here however, we work in the assumption of having a
precise measure of $c_v$: this is a reasonable and realistic
assumption, since there are several other processes which will allow
to probe $C_{WWH}$ independent from the process we are interested
in~\cite{Klute:2013cx}, and definitely with a shorter time scale with
respect to that needed to accumulate the luminosity required to
observe $tHj$.

As mentioned above, we are interested in the possibility to observe
and measure the $tHj$ cross section by looking at the $H\to
\gamma\gamma$ decay. Because this branching ratio depends on $C_{WWH}$
and $C_{ttH}$, the measurement of the total cross section with a
diphotonic final state cannot be straightforwardly translated into a
limit on the top Yukawa coupling. We consider two ways to deal with
this issue. The first possibility is to just rely on the fact that by
the time when this measurement will be possible, the LHC will have
completed a ``legacy'' measurement of the $H\to\gamma\gamma$
branching, which can be used as an input for our proposed
analysis. Alternatively, one can also include the dependence on the
$c_t$ factor entering in $H\to\gamma\gamma$, assuming that only the
top Yukawa is allowed to float (in which case the total Higgs width
stays approximately unchanged once $C_{WWH}$ is fixed, because the
dominant $H\to b\bar{b},c\bar{c},\tau\bar{\tau}$ and $VV$ partial
widths are fixed).  We will report two different confidence limits for
the Yukawa coupling: The first limit follows form a SM-like $H\to
\gamma\gamma$ branching ratio and the second one includes the
back-reaction of the modified top Yukawa coupling on the Higgs decay
phenomenology.

We will derive these constraints from characteristic angular
observables of the exclusive $tHj$ final state after showering,
hadronization and signal vs. background enhancing selection cuts.  In
our study we have identified several variables sensitive to the size
of $c_t$. For this paper, we have chosen $ R(H,
j_b)=\sqrt{\Delta\phi(H, j_b)^2+\Delta y (H, j_b)^2}$, {\it i.e.} the
distance between the $b$ jet and the reconstructed Higgs boson in the
$(y,\phi)$ plane as a single discriminating variable, since this
observable optimizes the discriminative power between different signal
hypothesis in the presence of realistic cuts, as we will show in the
next section. We will also discuss the sensitivity of $\Delta y (H,
j_b)$ to the value of $c_t$.

To understand the typical kinematics of the final state, we start by
reminding the reader that the cross section for $tHj$ production is
minimal for a SM-like top Yukawa value ($c_t=1$). This is due to
destructive interference between the diagrams of Fig.~\ref{fig:FD}
becoming maximal~\cite{Biswas:2012bd}. It is instructive to study some
leading-order parton-level distributions~\cite{Campbell:2013yla} in
presence of very generic cuts:
\begin{eqnarray}
  \label{eq:basiccuts}
  \mbox{lepton}&:&\ p_{T,\ell}\geq 10 \mbox{GeV},\ |\eta_\ell|<2.5\,,\nn\\
  \mbox{photons}&:&\ p_{T,\gamma}\geq 30 \mbox{GeV},\ |\eta_\gamma|<2.5\,,\
   R(\gamma ,\gamma)>0.1\,,\nn\\
  \mbox{jets}&:&\ p_{T,j}>20 \mbox{GeV},\ |\eta_j|<4.5\,.
\end{eqnarray}
Jets are obtained clustering the final state partons with
{\sc{FastJet}}~\cite{fastjet}, using the anti-$k_T$
algorithm~\cite{antikt} with $R=0.4$.

First of all we notice that the light jet associated to the light
quark current is typically produced at relatively small transverse
momentum ($p_{T,j}$ peaks at $\sim 40~$GeV) and high rapidity
$(|y_j|\sim 3)$. Hence cutting away events with central light jets
will not deplete significantly the signal, helping therefore in
enhancing the signal vs. background ratio. We also notice that
typically the top quark and the light jet lie in opposite hemispheres,
and as a consequence the heavy objects in the final state are
distributed such that the top quark is typically further away in
rapidity from the light jet than the Higgs boson. In the next section,
we will make use of these properties of the signal's kinematics to
design a cut flow that affects the signal rates as less as possible.

In the left panel of Fig.~\ref{fig:yt_dytH_LO} the leading-order
distribution of the reconstructed top (and anti-top) rapidity is
shown\footnote{Unless otherwise stated, all the following
  distributions are obtained by summing the cross sections for $tHj$
  and $\bar{t}Hj$ production.} for the two representative values of
$c_t$ that will be used in the following, after applying the cuts in
Eq.~\ref{eq:basiccuts}. Since we want to concentrate on shape
differences, here we show unit-normalized curves, but we stress that
the total cross-section for $c_t=0.5$ is a factor $\sim$ 1.5 larger
than the SM value. Together with this observation, the plot in the
left panel of Fig.~\ref{fig:yt_dytH_LO} shows that the interference
term is significantly bigger in size, and negative, when tops are
central: consequently it is clear that the smaller $c_t$ is, the more
central the top quarks are, and, conversely, tops are fairly uniformly
distributed in the central rapidity region when $c_t=1$. These
observations apply as well for the reconstructed Higgs (not shown): in
the SM scenario, $y_H$ is essentially flat for $y_H\in [-1,1]$,
whereas for ``BSM'' scenarios the Higgs rapidity tends to peak at 0.

In the right panel of Fig.~\ref{fig:yt_dytH_LO} we show how this
pattern translates to the rapidity distance between the top and the
Higgs, which is related to the observables in which eventually we will
be interested, {\it i.e.} distances between the reconstructed Higgs
and the hardest $b$-jet. In the SM case the negative interference
affects very sizeably the $|\Delta y(t, H)|\sim 1$ region, creating a
visible shape difference between the two signal hypothesis. As we will
observe in Sec.~\ref{sec:14tev}, the slope shown in the ratio panel on
the right persists even in presence of the other cuts that will be
introduced to enhance $S/B$, and affects also $R(H, j_b)$ and $\Delta
y(H, j_b)$ too, which we will use to set exclusion limits.

In anticipation of the main results, we also show how the $\Delta y
(t, H)$ distributions look when we split the total cross section by
requiring the $b$-jet to be harder (softer) than the light
jet. Fig.~\ref{fig:dytH_jb_bj_LO} shows that, when $p_{T,j_b} <
p_{T,j}$, the above picture is not qualitative changed. However, for
$p_{T,j_b} > p_{T,j}$, the Higgs boson and the top quark are much
closer in rapidity when $c_t=0.5$, as shown in the right panel of
Fig.~\ref{fig:dytH_jb_bj_LO}. This is due to the fact that to have
hard $b$-jets, the parent tops need to be more central: when $c_t=1$,
this situation is strongly disfavored by the negative interference, as
commented above, whereas a large part of the cross section in the
$c_t=0.5$ case is concentrated in this phase-space region, as shown in
the main panels.  In the next section we will use this very large
shape differences in $\Delta y$ distributions, in the regime where
$p_{T,j_b} > p_{T,j}$, as an extra-handle to set stronger constraints
on the top Yukawa.

\section{Prospective sensitivity and discovery thresholds at 14 TeV}
\label{sec:14tev}
In the following we perform a hadron-level analysis of the process
$pp\to (t\to \ell b \nu) (H\to \gamma \gamma) j$, $\ell=e,\mu$, at
14~TeV with a target luminosity of 3/ab. This will allow us give an
estimate of the discriminative power that is encoded in angular
observables after realistic selection criteria have been applied.

We investigate an exclusive final state that is obviously the cleanest
channel to observe $tHj$ production yet statistically limited due to
the small $H\to\gamma\gamma$ branching ratio. Side-band analysis
techniques are applicable and we can expect that systematic
uncertainties in this final state are small compared to multi
$b$-tagged events that have been discussed in the literature in the
context of parton-level
analyses~\cite{Farina:2012xp,Biswas:2012bd,Biswas:2013xva,Maltoni:2001hu}. We
include the following dominant irreducible and fake (jet faking
$b$-jets, jets faking photons\footnote{We use a flat factor of 1/1000
  for the jets faking photons~\cite{fakes}.})  backgrounds: $W^\pm
(H\to \gamma\gamma)$+jets, $W^\pm \gamma\gamma$+jets, $t
\gamma\gamma$+jets, $\bar t \gamma\gamma$+jets, $t\bar t
\gamma\gamma$+jets, $t\bar t (H\to \gamma\gamma)$+jets,
$t\gamma$+jets, and $\bar t\gamma$+jets.

All event samples are generated with {\sc{MadGraph}}~\cite{madgraph},
using the default {\sc{Cteq6l1}}\cite{Pumplin:2002vw} parton
densities, and are subsequently showered with
{\sc{Herwig++}}~\cite{herwig}. Quite obviously, a lot of systematic
limitations that are discussed in the context of $t\bar t H$ analyses
also impact this analysis, most notably the issues of heavy flavor
contributions that are still under investigation
presently~\cite{atlas}. The final state we consider in this section
will clearly minimize the sensitivity to these effects compared to
multi $b$-tagged final state, but heavy flavor production and tagging
still deserves a more detailed investigation in the context of $tHj$
production once the theoretical and experimental questions raised
in~\cite{atlas} are settled. A realistic in-depth analysis of the
corresponding uncertainties is currently not available and beyond the
scope of this section, therefore our results need to be understood
with a pinch of salt.

Our selection criteria closely follow the event topology that results
from the Feynman diagrams in Fig.~\ref{fig:FD}: we typically deal with
a central $b$ jet and a forward jet that are balanced by the
Higgs. Since we only have electromagnetic calorimetry in the central
part of the detector ($|\eta |<2.5$), we lose a fraction of the signal
due to the reconstruction in the central part of the detector. This is
unavoidable also for other decay modes, {\it e.g.}  for $H\to b\bar
b$, because $b$-jet identification relies on vertexing in the central
part of the detector too. We will continue to focus on $c_t=0.5$ for
illustration and also comment on $y_t>y_t^{\text{SM}}$ at a later
stage. The latter case is typically more complicate to constrain when
the back-reaction on $H\to \gamma\gamma$ is included.

\begin{figure*}[!t]
  \includegraphics[width=0.45\textwidth]{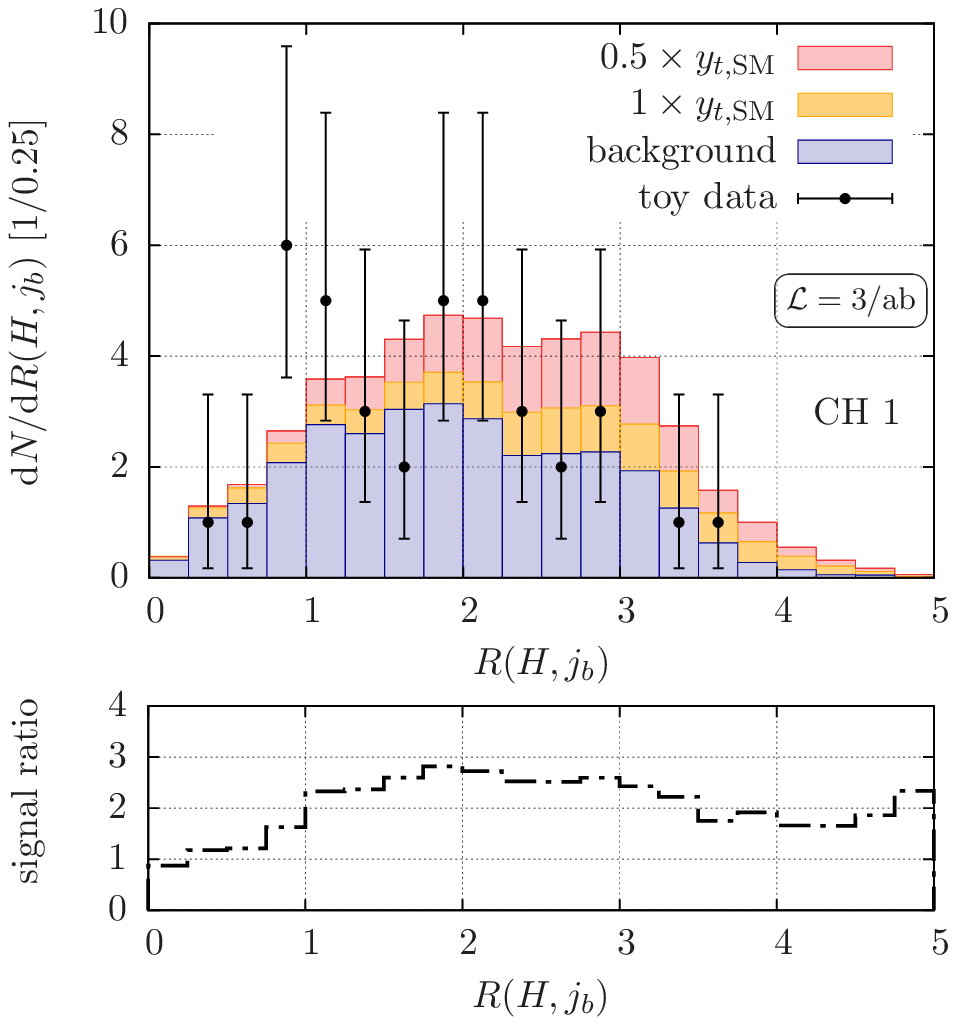}
  \hspace{0.05\textwidth}
  \includegraphics[width=0.45\textwidth]{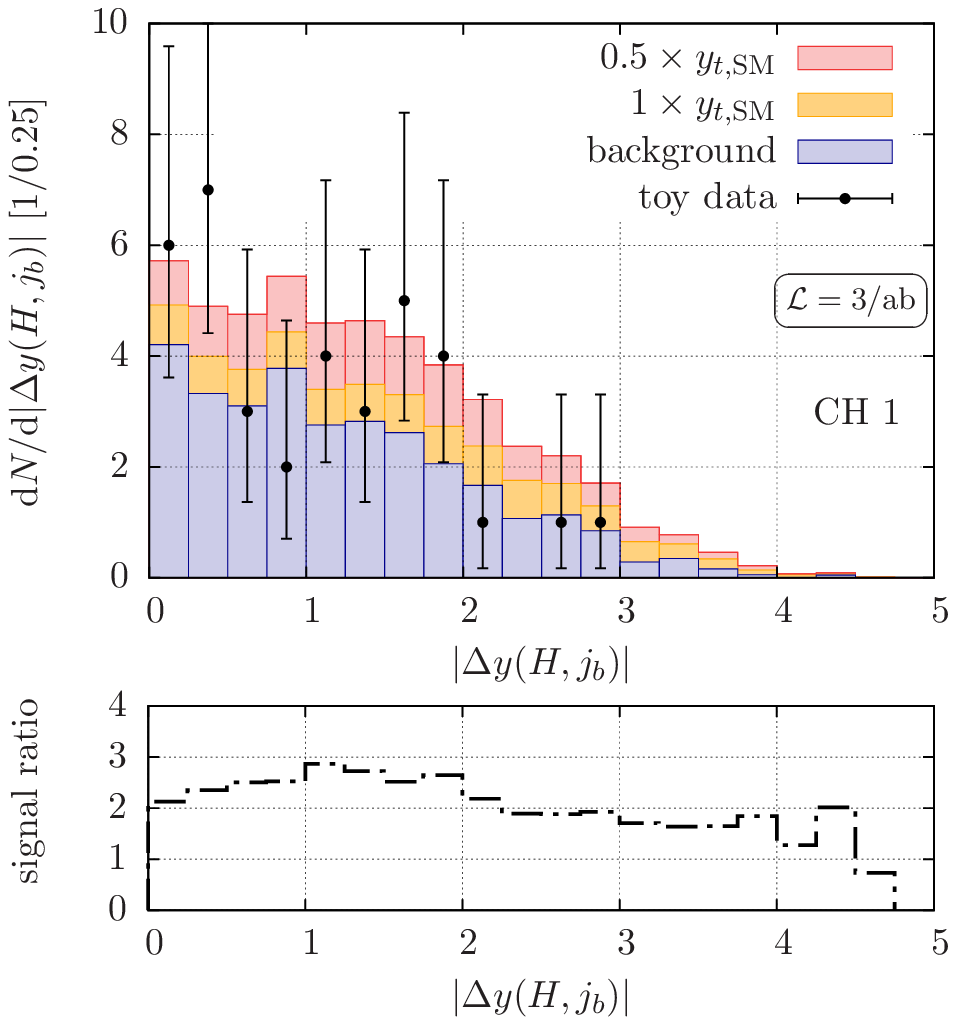}
  \caption{\label{fig:rbl} Lego-plot separation and rapidity
    difference of the reconstructed Higgs boson and $b$-tagged jet. We
    show the expected distribution for a target luminosity of 3/ab
    after the selection criteria detailed in the text have been
    applied. To get an idea of the involved statistical uncertainty of
    such a measurement with an SM-consistent outcome, we include toy
    data and the 95\% Bayesian confidence level error bars around the
    central values. We use these distributions and MC-sampled toy
    measurements to compute a confidence level interval for the top
    quark Yukawa coupling (see text); the $c_t=0.5$ sample includes a
    modified $h\to \gamma\gamma$ branching ratio. Note that the signal
    hypotheses overlap. The background does not contain the
    modifications due to $c_t<1$ for illustration purposes.}
\end{figure*}

In the actual analysis, we define isolated leptons and photons for
tracks which have less than 10\% energy deposit relative to the
tracks' transverse momentum in a cone of size $\Delta R=0.1$.
Leptons, photons and jets are then selected using the same basic cuts
reported in Eq.~\ref{eq:basiccuts}. We ask here for {\it exactly} one
lepton, two photons, and two jets (this reduces the reducible
backgrounds but also the signal).

The two photons need to be isolated $R(\gamma_1 ,\gamma_2)>0.1$ and
need to reproduce the Higgs mass of 125 GeV within
$|m(\gamma_1,\gamma_2)-125~\text{GeV}|<10~$GeV. The jets need to be
separated by $R (j_1,j_2)\geq 2$. Subsequently, we use a two
channel approach to formulate limits on the top Yukawa coupling,
described in the following:
\begin{enumerate}[1.)]
\item We order the jets in hardness $p_{T,j_1}>p_{T,j_2}$. $j_2$ needs
  to be central with $|\eta_{j_2}|<2.5$, and needs to pass a $b$-tag,
  whereas $j_1$ is in the forward region $|\eta_{j_1}|>1.0$. We use a
  working point with an efficiency of 85\% and a fake rate of
  10\%~\cite{btagging}. The isolated lepton and jet $j_2$ needs to
  have an invariant mass $m (j_2, \ell)<200~$GeV and $j_1$ needs to be
  separated from the lepton by $R (\ell, j_1)\geq 1$, and from the
  reconstructed Higgs ($p_H=p_{\gamma_1}+p_{\gamma_2}$) by $ R (H,
  j_1)\geq 2.5$.
\end{enumerate}
This cut flow is designed in such a way that we gain sensitivity to
$c_t=1$ over the background.\footnote{In particular the cuts on the
  lego-plot separations among the various objects help in reducing the
  backgrounds without depleting the signal too much: as we noticed in
  Sec.~\ref{sec:pheno}, the signal cross section is indeed
  characterized by ``large'' distances between the light jet and the
  other heavy objects.}
Specifically, in this ``SM region'' we expect ${\cal{O}}(10)$ signal
events (see Tab.~\ref{tab:xsecs}), that can in principle be used to
calibrate the measurement. We refer to this selection as ``Channel 1''
(CH1). The second selection is better tailored to BSM-induced effects,
yet statistically independent from CH1:
\begin{enumerate}[1.)]
  \setcounter{enumi}{1}
\item We order the jets $p_{T,j_2}>p_{T,j_1}$ and proceed subsequently
  exactly as described in 1.).  In particular, this amounts to events
  with harder $b$ jets, and invariant $j_2,\ell$ mass cuts on the
  harder jet.
\end{enumerate}
We refer to this selection as ``Channel 2'' (CH2).

After these steps we end up with cross sections of the two searches as
detailed in Tab.~\ref{tab:xsecs}.

\begin{table}[!b]
  \begin{tabular} { || c || c  |  c  |  r || }
    \hline
    \multirow{2}{*}{Channel 1} & \multirow{2}{*}{$\sigma_B=10.09~\ab$} & $\sigma_S=2.92~\ab$ &
    $c_t=1.0$ \\
    && $\sigma_S=4.80~\ab$ & $c_t=0.5$ \\
    \hline
    \multirow{2}{*}{Channel 2} & \multirow{2}{*}{$\sigma_B=6.3~\ab$} &
    $\sigma_S=2.02~\ab$ & 
    $c_t=1.0$ \\
    && $\sigma_S=4.42~\ab$ & $c_t=0.5$ \\
    \hline
  \end{tabular}
  \caption{\label{tab:xsecs} Signal and background cross sections as
    for the two selections as described in the text.}
\end{table}

\begin{figure*}[!t]
  \includegraphics[width=0.45\textwidth]{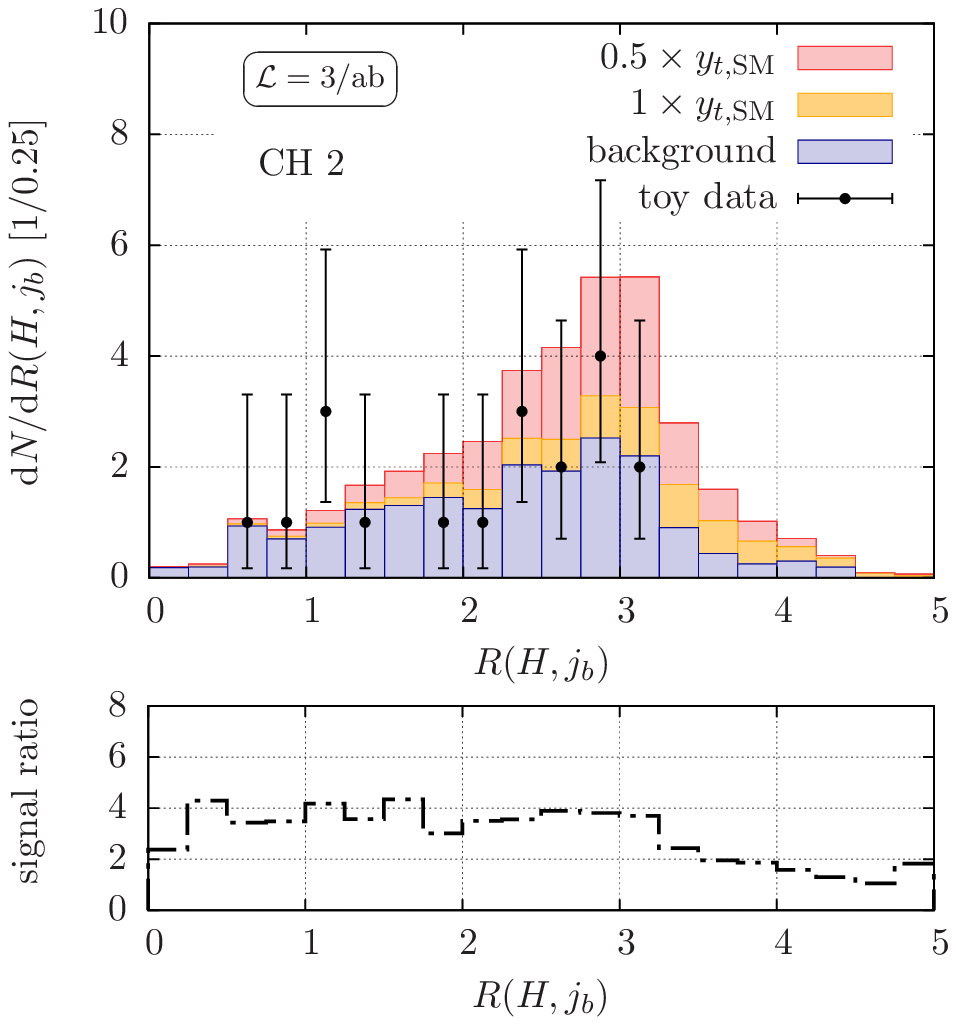}
  \hspace{0.05\textwidth}
  \includegraphics[width=0.45\textwidth]{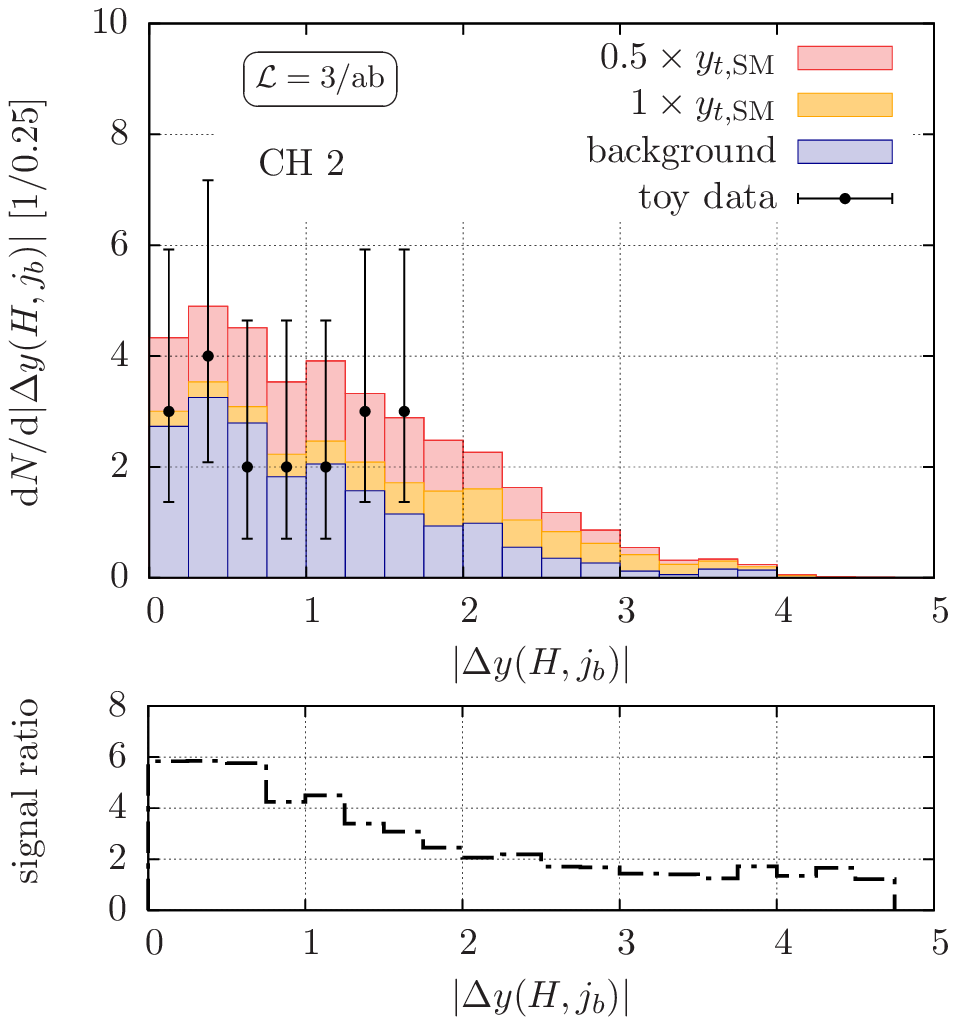}
  \caption{\label{fig:rbl2} Same as Fig.~\ref{fig:rbl} for a measurement
    performed in channel 2 as defined in the text.}
\end{figure*}

As explained previously, an appropriate choice of a discriminative
collider observable that encodes sensitivity to the top Yukawa
coupling is the rapidity difference between the reconstructed Higgs
and the $b$-jet. It also feeds into the lego-plot separation $R (H,
j_b)$ which, following our earlier discussion, is also sensitive to
the top quark Yukawa coupling, as can be seen in Figs.~\ref{fig:rbl}
and \ref{fig:rbl2}.

Note that our cut flow does not directly cut on this observable,
although the requirements $p_{T,j_b} \lessgtr p_{T,j}$ changes the
behaviour of $ R (H, j_b)$ and $\Delta y (H , j_b)$, as anticipated
in Sec.~\ref{sec:pheno}. As can be seen in Fig.~\ref{fig:rbl2}, in the
CH2 scenario, a better discrimination of $c_t$ can be achieved.

To compute a confidence level interval for the top Yukawa coupling we
perform a binned log-likelihood analysis~\cite{junk,cranmer} as
invoked by the experiments (see {\it
  e.g.}~\cite{newboundsb,newboundsa,Englert:2012cb} for details and
validation) on the basis of the distributions of
Fig.~\ref{fig:rbl}. We use the CLs method~\cite{CLS} to formulate a
lower limit on the top Yukawa interactions. Since the top Yukawa
coupling interferes destructively with the remaining contributions in
Fig.~\ref{fig:FD}, we obtain a larger cross section for
$y_t<y_t^{\text{SM}}$ for fixed top and Higgs masses and widths and
can formulate constraints.

Scanning over different signal event samples with varied $y_t$,
keeping track of the differential cross section modifications, we
compute a lower limit (keeping $C_{WWH}=1$)
\begin{equation}
  c_t\gtrsim 0.5~\hbox{at 67\% CLs~[80\% CLs]}\,,
\end{equation}
where the number in brackets refers to the confidence level when the
modified $y_t$ feeds into $h\to \gamma\gamma$ in the signal sample
(for comparison reasons we take Fig.~\ref{fig:rbl2} at face
value). The differential cross section indeed contains valuable
information which is not accessible by only counting events: the
confidence level for a CLs test based on total event counts we only
exclude $c_t \lesssim 0.5$ at 58\% CLs~[73\% CLs].

We now include the information of channel 2 to the picture. In this
case, as we have explained above in detail, the ratio between $\Delta
R$ distributions has a different (inverted) shape at small values and
provides increased statistical pull. Despite that in this case $S/B$
is not as optimal for the SM scenario as before, we add statistical
information that efficiently constrains $c_t$ across the two
regions. We end up with confidence levels for our benchmark point
(modifications of the Higgs-included background contributions are
taken into account)
\begin{equation}
  c_t\gtrsim 0.5~\hbox{at 95\% CLs~[99\% CLs]}\,.
\end{equation}
Similarly we can try to formulate an upper bound for $y_t$. The $tHj$
production cross section starts to grow for $y_t > y_t^{\text{SM}}$,
but the $H\to \gamma \gamma$ branching ratio falls quickly and stays
small because of a increasingly preferred gluon-phillic Higgs
decay. This leads to a much looser constraint when we include the
back-reaction of the modified top Yukawa coupling on the diphoton
branching. We obtain
\begin{equation}
  c_t\lesssim 1.6~\hbox{at 95\% CLs~[85\% CLs]}\,,
\end{equation}
where the number in brackets corresponds again to the constraint with
modified $H\to \gamma \gamma$.

\section{Summary}
\label{sec:conc}
The late discovery of the Higgs boson provides us a unique opportunity
to put the SM hypothesis to the ultimate test: Is the Higgs boson
really the one predicted by the SM or is it the harbinger of physics
beyond the SM? Theoretical prejudice based on TeV scale naturalness
inevitably forces the latter interpretation upon us. Given that the
top quark is of crucial importance for natural TeV scale due to its
large Yukawa interaction $y_t$, the top-Higgs sector is a
well-motivated playground to look for deviations from the SM
expectation.

In this paper, we have argued that we should be able to constrain the
Yukawa coupling at the high luminosity LHC (3/ab) at 14~TeV, even if
we focus on rare final states of $tHj$ production.  Tree-level
destructive interference effects steered by $y_t$ result in modified
angular correlations and signal cross sections motivate measurements
based on angular correlation-inspired collider observables as
well-adapted search strategies for deviations from the SM. To maximize
the sensitivity to $y_t\neq y_t^{\text{SM}}$, we employ an analysis
approach that is based on two complementary selections of the
exclusive rare final states that results from leptonic top decays and
$H\to \gamma\gamma$. The first one is a ``traditional'' signal
vs. background discrimination that adapts to the SM expectation and
seeks to gain as much signal events as possible in the SM-context (and
calibrating the measurement). The second complementary selection
adapts to a phase space region that is mostly dominated by $y_t\neq
y_t^{\text{SM}}$-induced modifications of the showered differential
angular observables. Combining the two selections we have shown in
detail that $y_t\lesssim 0.5\, y_t^{\text{SM}}$ can be excluded at
95\%-99\% CL and similarly $y_t\gtrsim 1.6\, y_t^{\text{SM}}$ at
85\%-95\% CL, already in this channel (depending on the assumptions
made on the $H\to\gamma \gamma$ branching). Since we do not rely on
any details of the Higgs system, our approach can straightforwardly be
generalized to other Higgs decay modes.

\acknowledgments 
CE thanks B.~Aacharya, J.~Ferrando, M.~Takeuchi, and M.~Spannowsky for
helpful conversations. ER is grateful to U.~Haisch for giving an
initial motivation to look into
Refs.~\cite{Biswas:2012bd,Farina:2012xp}, and for several interesting
discussions, and to B.~Mele for an useful and encouraging discussion
in Trento. CE is supported in parts by the IPPP Associateship
programme.


\end{document}